\documentstyle[12pt,aaspp4,psfig]{article} 

\def\mathnew{\mathsurround=0pt}
\def\simov#1#2{\lower .5pt\vbox{\baselineskip0pt \lineskip-.5pt
        \ialign{$\mathnew#1\hfil##\hfil$\crcr#2\crcr\sim\crcr}}}
\def\simgreat{\mathrel{\mathpalette\simov >}}

\def\simless{\mathrel{\mathpalette\simov <}}

\def\g{{$\gamma$}}

\begin{document}
\title{ 
Simultaneous X-Ray and Gamma-Ray Observations of TeV Blazars:
Testing  Synchro-Compton Emission Models and
Probing the Infrared Extragalactic Background }
\author{Paolo S. Coppi}
\affil{Department of Astronomy, Yale University, P.O.
Box 208101, New Haven, CT 06520-8101}
\authoremail{coppi@astro.yale.edu}
\author{Felix A. Aharonian}
\affil{ Max-Planck-Institut f\"ur Kernphysik, Postfach 103980, D-69029 
Heidelberg, Germany}
\authoremail{aharon@fel.mpi-hd.mpg.de}
\centerline{\it to appear in ApJ Lett., vol. 521, L33}

\begin{abstract} 
The last years have seen a revolution in ground-based 
\g-ray detectors. We can now detect the 
spectra of nearby TeV blazars like Mrk 421 and
501 out to $\sim 20$ TeV, and during the strongest 
flares, we can now follow fluctuations
in these spectra on timescales  close to the shortest
ones likely in these objects.  
We point out that this represents
a unique opportunity. Using these and future detectors
in combination with broadband X-ray satellites like SAX and RXTE, 
we will be able to simultaneously follow all significant 
X-ray/\g-ray variations in a blazar's emission. This
will provide the most stringent test yet of the synchrotron-Compton
 emission model for these objects. In preparation for the data to come,
we present sample SSC model calculations using 
a fully self-consistent, accurate code to illustrate
the variability behavior one might see
and
to show how good timing information can probe physical
conditions in the source. If the model works, i.e., if  X-ray/TeV
variations are consistent with being produced by a common 
electron distribution, then we show it is possible to 
robustly estimate the blazar's intrinsic TeV spectrum from
its X-ray spectrum. Knowing this spectrum, we can then
determine the level of absorption in the observed spectrum.
Constraining this absorption, due to \g-ray pair production
on diffuse radiation,  provides an important
constraint on the infrared extragalactic background intensity.
Without the intrinsic spectrum, we show that detecting
absorption is very difficult and argue that 
Mrk 421 and 501, as close as they are, may already be absorbed by a
factor 2 at $\sim$ 3 TeV. This should not be ignored when
fitting emission models to the spectra of these
objects.

\end{abstract}

\section{Introduction}

Recent broadband
spectral compilations (e.g., see Ulrich, Maraschi, \& Urry 1997)
suggest that blazars can be divided into two rough classes,
``LBL'' and ``HBL,'' depending on the energy where their
synchrotron emission peaks (optical/UV for LBL, and UV/X-ray for HBL).
As expected in a synchrotron-Compton (SC) jet emission model 
 where
the synchrotron-emitting electrons also Compton upscatter 
photons to \g-ray energies
(see, e.g.,  Sikora 1997 and Coppi 1997 for reviews of current
blazar emission models and controversies),
the energy of their   \g-ray emission peak (MeV/GeV in LBLs,
GeV/TeV in HBLs) is well-correlated with that of 
the synchrotron peak.  Most work to date
has focused on  LBL objects like 3C 279,
probably because these objects are the ones associated with
the most powerful, classical radio sources and 
also simply because they were  the only ones accessible in the \g-ray domain
(via  EGRET observations). In this Letter,
we argue that equal if not more effort should now be devoted to
the weak and ``uninteresting'' nearby HBLs like Mrk 421 and 501;
the scientific payoff is potentially much larger.
The new development is the arrival of powerful, 
ground-based Cherenkov telescope arrays (e.g., 
see Aharonian \& Akerlof 1997).
Such detectors have  enormous collection areas 
(up to $10^9\ {\rm cm}^2$ vs. $1500\ {\rm cm}^2$ for EGRET), and 
the already existing HEGRA array can  follow  
$\sim$ 500 GeV- 10 TeV spectral variations in 
Mrk 421/501 down
to $\sim$ one hour timescales (e.g., see Aharonian et al. 1999a). 
The sensitivity and energy coverage will significantly improve
once new and larger  arrays like  VERITAS and HESS come online.
This range of timescales and spectral coverage is well-matched
to the capabilities of modern broadband X-ray detectors like
ASCA, RXTE, and SAX that can monitor fluctuations in the 
synchrotron emission between $\sim$ 1-200 keV. Using HBLs, for the
first time we can thus  hope to follow and resolve simultaneously 
${\it all}$ significant fluctuations in the putative synchrotron and
Compton emission components of a blazar.  

This has two important consequences. First,  matching the observed 
X-ray/TeV lightcurves (as opposed to simply fitting
snapshot spectra obtained many days apart) 
provides a very stringent test of the SC emission model
since we have
{\it two} detailed handles on the {\it single} electron 
distribution responsible for both emission components.
The test can rule out alternate ``hadronic''models like that 
of Dar \& Laor (1997), where decaying pions produce an 
additional, non-Compton \g-ray component. 
Second,  if the SC model works during a large flare  (where the emission
from a single region may dominate), then we can try to use
the observed  X-ray spectrum  to predict 
the  TeV spectrum. This is key since
even for nearby HBLs like Mrk 421/501, 
the TeV photons propagating towards us 
can pair produce on the low
energy diffuse extragalactic background radiation (DEBRA) 
and thus be absorbed (Nikishov 1962).
If this absorption is present, the usual SC
X-ray/TeV spectral modeling of snapshot 
spectra (e.g., Mastichiadis \& Kirk 1997; Pian et al. 1998)
will fail. Note that by
using only the  relative time behavior of the X-ray 
and TeV fluxes to test the SC model, 
we avoid this problem. With an estimate for the intrinsic SC
\g-ray spectrum, then, and {\it only} then,
can we attempt to measure and correct for any absorption.
This last possibility is particularly exciting since 
a measurement constrains the 
density of the target DEBRA photons, which has
implications for galaxy evolution and cosmology 
(e.g., MacMinn \& Primack 1996).

Observational advantages aside, objects like Mrk 421 and 501 
are still better candidates. 
They have subluminous accretion disks, implying
a weak radiation field outside the jet. For large flares, a
simpler and hence more constraining  SSC (synchrotron self-Compton) 
emission model  may thus be sufficient. 
Even if external photons are important, HBLs give
tighter constraints because the Compton
scatterings responsible for the TeV \g-rays are probably
in the Klein-Nishina limit where the exact target photon
energies do not matter. 
Finally,  nearby sources that emit to $\simgreat 20$ TeV 
 allow us to probe the DEBRA at  10-30 $\mu\rm m,$ 
the range most difficult to constrain via other techniques. 
In sum, 
simultaneous X-ray and TeV observations of nearby HBLs
can tell us much about AGN jet emission mechanisms and the
level of the infrared/optical (IR/O) DEBRA --   but only if 
{\it both} aspects of
the problem (the emission and the absorption) are attacked
concurrently.  In \S 2 below, we review
how \g-ray spectra are modified by absorption and show
that its effects may be important even in Mkn 421/501.
In \S 3, we show  examples of the rich 
spectral variability that even simple SSC models can produce
as well as useful timing diagnostics to probe it.
If an SC model works, we demonstrate how to 
robustly estimate the shape of the TeV spectrum
using  X-ray data.  We conclude in \S 4.

\section{ Gamma-Ray Pair Production on Diffuse Background Radiation }

The interaction of GeV/TeV radiation with the IR/O DEBRA
has received considerable attention, e.g., see 
Madau \& Phinney (1996), MacMinn \& Primack (1996), 
Coppi \& Aharonian (1997), Biller et al.
(1998),  Stanev \& Franceschini
(1998),  and Stecker \& DeJager (1998) for
some of the recent papers. A detailed
discussion of the relevant transfer equations can be found there.
In general, to obtain
the mean free path for a \g-ray of energy $E_\gamma,$ 
one must convolve the DEBRA photon number distribution,
$n(\epsilon),$  with the pair production cross-section
(e.g., see Gould \& Schr\'eder 1966).  However, this cross-section is 
peaked and for nearby ($z\ll1$) HBLs
and almost all plausible DEBRA shapes,  
over half the interactions occur on
DEBRA target photons
with energies $\epsilon = 0.5-1.5 \epsilon_\ast,$ where
$\epsilon_\ast =
4m_e^2c^4/E_\gamma\approx 1.04 (E_\gamma/1{\rm TeV})^{-1}{\rm\  eV}.$ 
To accuracy better than $\sim 40\%,$
we can thus approximate
the absorption optical depth as
$$\tau_{\gamma\gamma}(E_\gamma) \approx 0.24 ({E_\gamma
\over 1 TeV})({u(\epsilon_\ast) \over
10^{-3} {\rm eV} {\rm cm}^{-3}})({z_s \over 0.1}) h_{60}^{-1}.$$
Here $u(\epsilon_\ast)=\epsilon_\ast^2
n(\epsilon_\ast)$ is the typical energy density in a
energy band centered on $\epsilon_\ast,$  $h_{60}$ is
the Hubble constant in units of $60 {\rm km}{\rm s}^{-1}
{\rm Mpc}^{-1},$ and $z_s$ is the source redshift.
If $I_0 (E_\gamma)$ is
the intrinsic source spectrum,  the corresponding observed spectrum
is then $I(E_\gamma)=I_0(E_\gamma)\exp(-\tau_{\gamma\gamma}).$ 
Note that if the DEBRA spectrum near $\epsilon_{\ast}$
can be approximated by a power law, $n(\epsilon) 
\propto \epsilon^{-\alpha_\ast},$ then $\tau_{\gamma\gamma}$
at energies $E\sim E_\gamma$ goes as $E^{\alpha_{\ast}-1}.$
Connecting
the COBE far IR  measurements to the latest UV background
estimates, one gets a crude DEBRA spectral index $\alpha \sim 2$ 
(e.g., see Dwek et al. 1998 for a good compilation of the latest DEBRA
 observations and models). To zeroth order, then, $\tau_{\gamma\gamma} 
\propto
E_\gamma,$ and the observed spectrum
should  be $\sim I_0(E_\gamma)\exp(-E_\gamma/E_c)$ where 
the cutoff energy $E_c$ is set by
$\tau_{\gamma\gamma}(E_c)=1.$ Interestingly, this is exactly
the type of shape seen in Mrk 501 by Whipple (Samuelson et al. 1998) and
especially by HEGRA, which measured the spectrum in the exponential
tail up to energies $\sim$ 20 TeV 
(Aharonian et al. 1999b). Does this mean we are seeing 
absorption? No. The inset in the
upper right corner of Fig. 1 shows a Mrk 501-like SSC 
spectrum in the TeV energy region. The dotted and dashed 
curves respectively show the absorbed spectra for 
an IR/O DEBRA level at the low end of estimates  
($u_{\rm l} = 2\times 10^{-4} \, {\rm eV}
{\rm cm}^{-3}{\rm s}^{-1}$)  and at the  high end
 ($u_{\rm h} = 2\times 10^{-3} \, {\rm eV}
{\rm cm}^{-3}{\rm s}^{-1}$) assuming a source
redshift $z_s$ and taking $\alpha=2.$ 
The absorbed spectra look just 
like unabsorbed spectra from blazars with 
lower cutoffs in their electron energy
distributions.  To next order, 
the  DEBRA is  better described 
as the sum of two  
emission components 
( starlight from galaxies peaking at $\sim 1$ eV, and
dust re-emission peaking at $\sim 100 \mu\rm m$). In models,
the $ 1-10 \mu\rm m$ side
of the ``valley'' between the DEBRA emission peaks is typically 
a power law with  $\alpha \sim 1.$ At the 
corresponding $E_\gamma\sim 1-10$ TeV, roughly the energy range
of current TeV detectors,
$\tau_{\gamma\gamma}$ is thus almost constant!
The shape of the spectrum is unchanged and
again we cannot infer absorption. Note that  
recent results at $\epsilon_\ast \sim 3 \mu\rm m$
(Dwek \& Arendt 1998) give a high DEBRA energy density,
$u(3\mu\rm m)\sim 2\times 10^{-3}\, {\rm eV}
{\rm cm}^{-3}{\rm s}^{-1}.$ 
Even for Mkn 501
($z_s = 0.03$),
$\tau_{\gamma\gamma} \approx 0.5$ at $E_\gamma \sim 3$
TeV, i.e., absorption corrections  may be important
($I_0/I \sim 2)$! For a $z_s\sim 0.1$ like 
that of PKS 2155-304 (a possible TeV source, see Chadwick
et al. 1999), they are almost certainly important.
The strongest DEBRA constraints may in fact
come from energies $E_\gamma \sim 10-30$ TeV, which
probe DEBRA energies on the ``other''
side of the valley ($\epsilon_\ast\sim 5-60\mu\rm m,$). Here, 
$\alpha > 2$ and absorption should
grow  {\it super-exponentially} with \g-ray
energy. To resolve such sharp absorption cutoffs
 requires good energy resolution, 
$\Delta E/E \simless 20\%,$ which 
appears achievable by Cherenkov arrrays
operating in stereoscopic mode (Aharonian et al. 1997).

\section{SC Emission Models and the Intrinsic Gamma-Ray Spectra of Blazars} 

To test SC models one first requires a 
theoretically accurate and realistic emission model.
Although there are many  SC calculations in the literature, we caution
that many do not apply to HBLs such as Mrk 421/501
where much of the Compton scattering is 
probably in the Klein-Nishina regime.
The key differences in this regime are that
an electron scattering off  low energy 
photons loses essentially {\it all} its energy to 
the photon and  Compton cooling becomes inefficient compared to 
synchrotron cooling. This changes, for example, the
mapping between the synchrotron and \g-ray emission
components.  In HBLs, the
peaks of the synchrotron and Compton emission components are 
{\it not}
produced by electrons of the same energy.
Also, the Compton (\g-ray) flux at the highest energies will tend to track the 
X-ray synchrotron flux only {\it linearly}
(e.g., see Ghisellini, Maraschi, \& Dondi 1996),
instead of quadratically as expected in the Thomson case.
Approximations such  as
using a Klein-Nishina ``cutoff''
for the Compton cross-section, or solving for the electron 
energy distribution using a continuous energy loss approximation
are dangerous when applied to HBL objects and can
easily lead to quantitative errors of factors of several
in the spectra  predicted for the peaks and tails
of the Compton and synchrotron components 
(e.g., Coppi \& Blandford 1990, Coppi 1992). 

To illustrate the wide range of time behavior possible in even
a  simple, one-zone, homogeneous SSC model,
we present
calculations using the code of Coppi (1992). This code
is fully self-consistent and uses no approximations
for the emission processes. The calculations in Fig. 1-3
use basic parameters which reasonably describe the
time behavior of Mrk 421 (Mastichiadis \& Kirk 1997),
except we increase the electron cutoff energy
by a factor 30 to model Mrk 501 (which has higher energy
synchrotron emission, better suited for our purposes).
 Figure 2 shows
 the model's response to three types of variations:
(Fig. 2a) the electron injection compactness $l_e$ 
($=L_e \sigma_T/R m_ec^3$, where $R$ and $L_e$ 
are respectively the rest frame source radius and 
electron luminosity) varies randomly by 
$\pm 50\%$ every light
crossing time ($R/c$) 
with all other parameters kept fixed, (Fig. 2b) $l_e$ varies
as in Fig. 2a but now the electron cutoff energy also varies
with $l_e$ ($\gamma_0 \propto l_e$), and 
(Fig. 2c) $l_e$ varies in Fig. 2a but now
the magnetic field varies with $l_e$ ($B \propto l_e$).
Note that for rapid, small variations, the Compton \g-ray flux tracks the 
synchrotron X-ray flux linearly, i.e., not in the standard quadratic
manner. In the Klein-Nishina limit, the main target photons 
for producing TeV \g-rays are   IR/O synchrotron photons.
These are produced by low energy electrons with very long cooling times
($> R/c$) that cannot respond quickly to rapid ($\sim R/c$)
changes in the electron injection ($l_e$). If one cannot directly
observe the IR/O emission (which in real sources may dominated
by emission from other parts of the jet), this means
it is {\it not} obvious what  target photon distribution to use
when fitting observations.
Note that in general, the responses in Fig. 2a-c look quite
different, i.e., good timing studies are a powerful diagnostic.
With well-sampled lightcurves,
we can make (DEBRA absorption independent!) 
 cross-correlation diagrams such as  Fig. 3.
At keV energies, in curves (a) and (b) of Fig. 3, 
we see soft-hard lag behavior (due
to the finite electron cooling times) similar to that observed in
Takahashi et al. (1996). Such behavior is washed out, though, 
when the magnetic field  changes significantly during a flare
(curve c) and the mapping between observed synchrotron 
photon energy and emitting electron energy is destroyed.
At $\sim 100$ keV, we also clearly 
see the transition from
synchrotron emission to Compton emission: the lag jumps up
suddenly because the Compton radiation comes from  {\it lower}
energy electrons.  The \g-ray to  X-ray lag 
decreases with increasing energy, but does {\it not} go to zero.
In an SSC model, it takes at least $\sim R/c$  to 
significantly change the target photon intensity in the source.
If the target photons are external and do not vary, no such lag
should be seen. 

Figure 2 also shows behavior that
allows us to make  robust predictions for HBL \g-ray
spectra. Note the \g-ray hardness ratios in Fig. 2a and 2c
fluctuate very little. When 
the cooling of energetic  electrons is dominated 
by synchotron radiation, the {\it only} way to change the shape 
of the cooled electron distribution
is to change the shape of the electron 
injection function (e.g., as in Fig. 2b).  Since 
the TeV gamma-ray spectrum in this case is essentially 
the cooled TeV
 electron distribution, the shape of the \g-ray spectrum is
insensitive to most source parameters.  In 
particular, it does not depend strongly on the target photon
distribution, as shown in Fig. 1, where a completely different
distribution gives the same Compton spectrum 
(long-dashed curved in the inset).
If we  can ``invert'' the observed synchrotron X-ray spectrum
to obtain the underlying electron distribution
(e.g., as in Fig. 1), we have  all we need to
predict the shape of the upscattered TeV spectrum. 
Extrapolating
from the  spectrum observed at low energies
where intergalactic absorption should  not be
important (e.g., 700 GeV in Fig. 1), we can 
then  predict the unabsorbed flux at TeV energies.
Our accuracy  is limited
by  uncertainties in $B_0$, $\delta,$ and the possible
presence of external,  low energy IR target photons (with too many
such photons, an electron does not lose most of its energy
in a typical Compton scattering).
However, bad estimates of $B_0$ and $\delta$ only
cause an overall energy
shift of the predicted \g-ray spectrum by a factor 
$({\delta \over B})^{1/2}$ (see Fig. 1),  i.e., a fairly weak dependence.
Also, the rest frame energy density of external IR photons must 
exceed  the synchrotron photon energy density to cause
significant deviations in the predicted spectrum. 
This is possible, but not likely in HBLs.
The inset of Fig. 1 shows the maximum error we could
produce  by playing with $\delta,$
$B,$ and $\lambda_s^l$ (which controls the number of 
low-energy target photons)
while requiring the prediction to 
match the observed  $200-700$
GeV spectrum. This corresponded to about a factor three uncertainty
at 10 TeV -- not bad considering our minimal  assumptions
and that typical DEBRA models
predict strong absorption at
such energies.

\section{Conclusions}

Mrk 421/501 and similar HBL sources provide ideal
laboratories to test in detail the emission models for these
objects. If we can show that
a simple SC model works during at least the strongest flares,
then we can use good broadband X-ray spectra of these sources to 
infer their intrinsic TeV spectra.
Then, and only then, can we look for evidence of \g-ray
absorption and attempt to constrain the IR/O DEBRA.
(Blazar modelers should  not forget that the Compton
spectra they are trying to fit could be strongly attenuated!)
In any single observation, the  absorption 
might be due both to intrinsic blazar
IR/O photons as well as intergalactic ones.
While these contributions can be difficult to disentangle,
Mrk 421 and 501 conveniently have the
same redshift. Thus, we can require that any
absorption attributed to
intergalactic photons  be exactly the same
for {\it all} flares in {\it both} sources. 
These two sources alone may  give us
the first firm handle on DEBRA \g-ray absorption.

\section{Acknowledgments} 
\noindent PSC was supported by NASA grant NAG 5-3686 and thanks
the Max-Planck-Institut
f\"ur Kernphysik for its generous hospitality.

\newpage
\figcaption[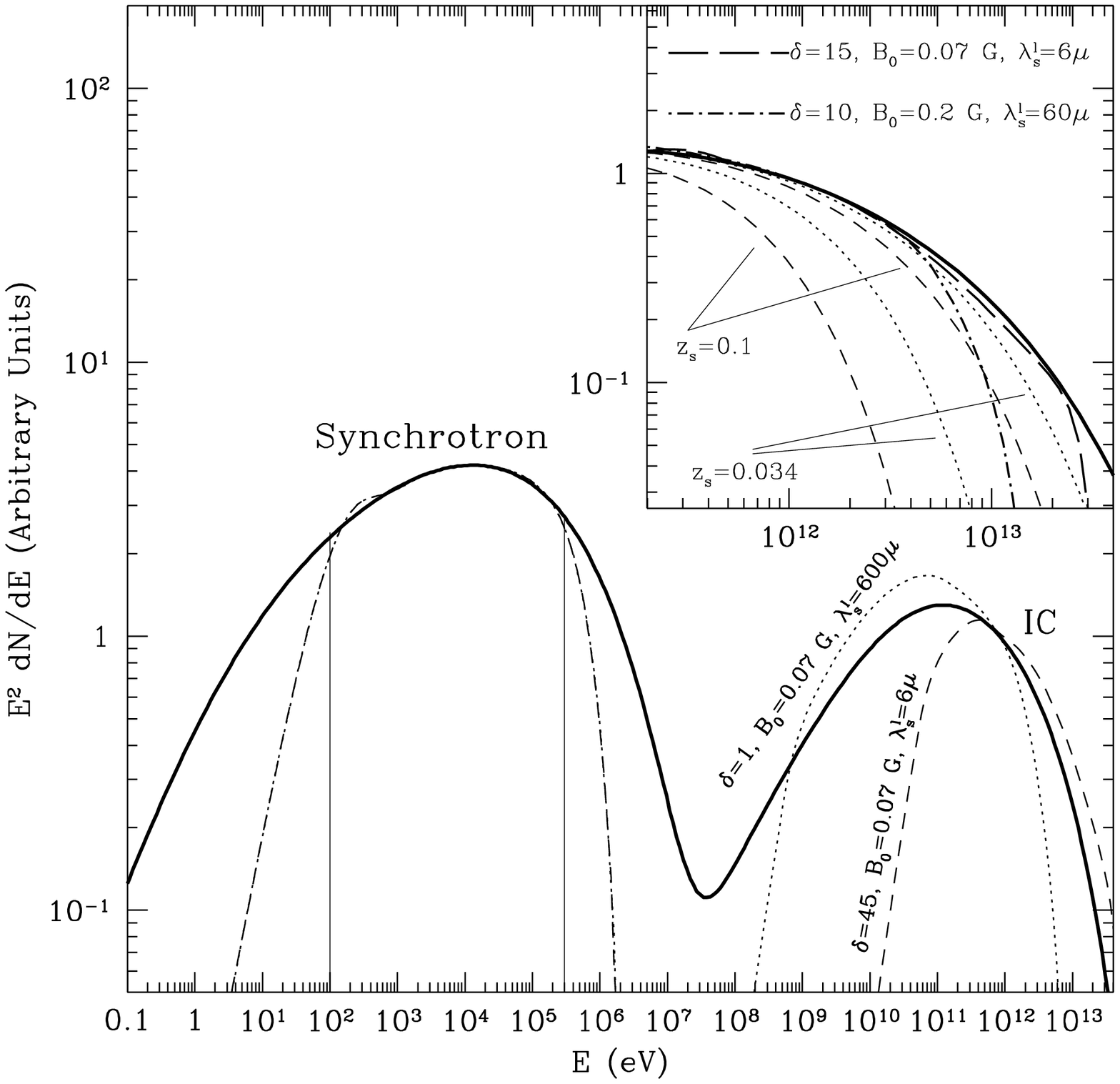]{
The SSC spectrum ({\it heavy solid lines}) produced by
a spherical source which has rest frame radius
$R = 4.7\times 10^{16}$ cm, is observed with Doppler boost
factor $\delta=15,$ and has a 
 tangled magnetic field $B_0 = 0.07$ Gauss.
Electrons enter  
with an initial (injection) energy spectrum $d \dot
N_{inject}(\gamma)/ d\gamma \propto (\gamma)^{-s} 
\exp(-\gamma/\gamma_0)$ down to   $\gamma_{min}=1.3,$
where $\gamma$ is the rest frame electron Lorentz 
factor,  $s=1.7,$ and  $\gamma_0=7\times 10^6.$  They escape
on a rest frame timescale $t_{esc} = 3.3 R/c.$
The spectrum shown is {\it not} a
steady-state, equilibrium spectrum, but rather
the integrated flux (from $t=0$ to $50 R\delta^{-1}/c$)
of the time-varying model in Fig. 2a.
The {\it dot-dashed} curve is the synchrotron emission from 
an electron distribution reconstructed from the 0.1-300 keV 
X-ray spectrum. (The distribution is obtained by
using a delta function  emission approximation to
initially ``invert'' the X-ray spectrum  and then iterating
using the exact synchrotron emission spectrum.)
The {\it dotted} and {\it dashed} curves in the main figure give the
Compton \g-ray  spectra predicted from the electron distributions
reconstructed assuming two extreme sets of  model parameters.
The target photon distribution used was not the synchrotron
spectrum, but a power law $n(\lambda) \propto \lambda^2$
extending (in the source frame) from 0.1 to $\lambda_s^l$ microns.
The inset in the upper right 
corner of the figure shows a blowup of the spectrum in the TeV
energy region. The {\it heavy dot-dashed} and {\it heavy long-dashed}
curves show the predicted gamma-ray spectra for two
more realistic sets of parameters.  The {\it dotted} and
{\it dashed}
 curves give the spectra obtained from a source described
by the solid heavy curve after \g-ray absorption by 
IR/O DEBRAs with energy densities $u_{\rm l}$ and $u_{\rm h},$
respectively (see text).
\label{fig1}}

\figcaption[fig2.ps]{
The SSC response to variations in electron injection and
magnetic field strength (see text). The top of each panel shows
the hardness ratios between various (observed) energy 
bands as a function of observer time. The bottom 
shows the photon number flux escaping in several (observed) energy bands.
The {\it heavy solid} curve is the total electron injection 
luminosity ($l_e(t)$) as a function of time (see text) 
and is the same for all three panels. 
At $t=0,$ the source is assumed to be 
empty and electron injection is turned on impulsively. 
The initial value for the total electron compactness (injection luminosity) is 
$l_e=2.4\times 10^{-4}.$ The flux levels shown 
are normalized to be the same 
at $t\approx 10 R\delta^{-1}/c.$ The hardness ratio normalizations
have also been adjusted for clarity.
Note the flux and hardness
ratio axes are logarithmic. The electron escape time is 
$3.3 R/c$ for the top and bottom panels, and $333 R/c$ for the 
middle panel.
\label{fig2}}
\figcaption[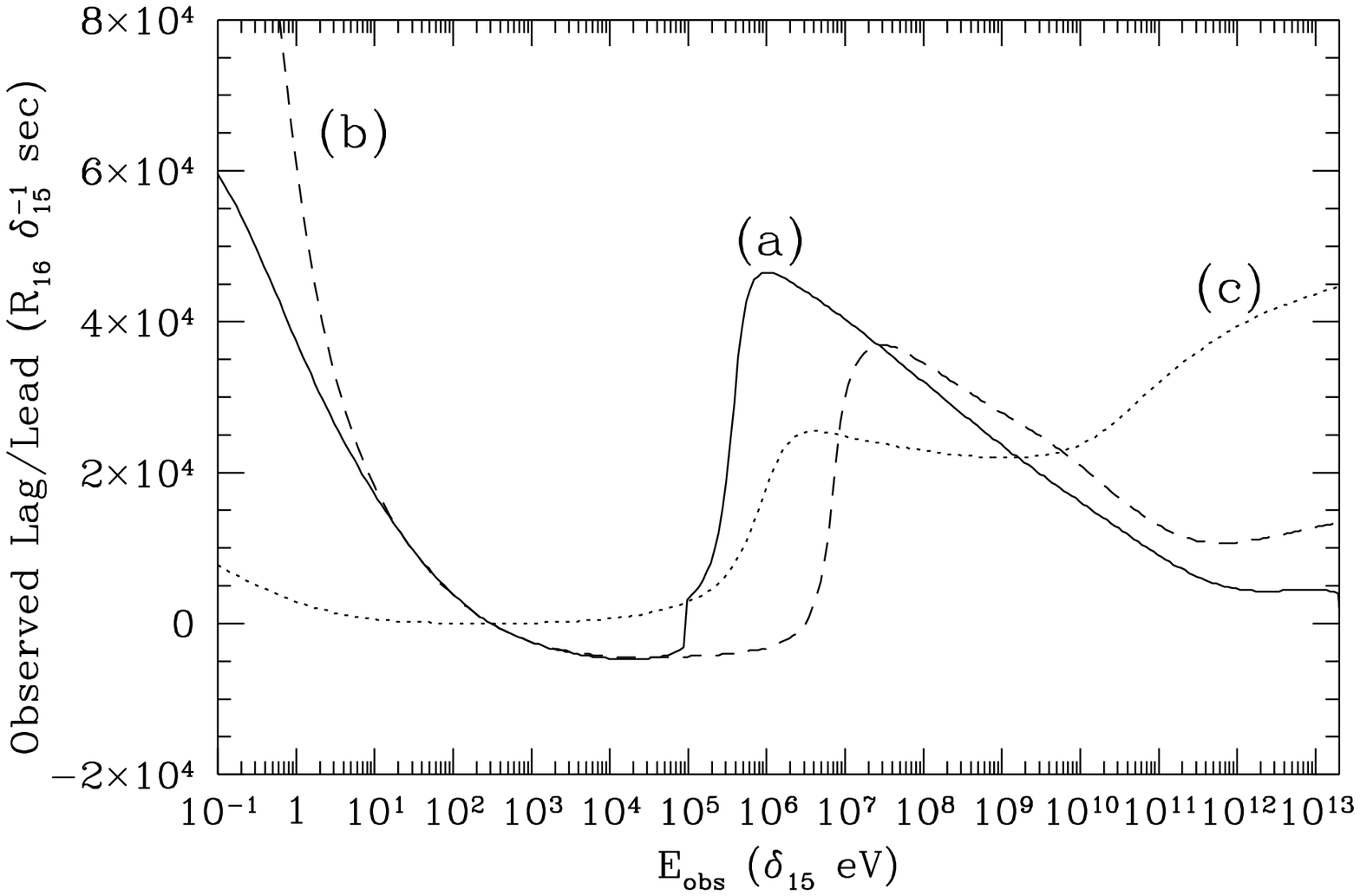]{ The relative lag/lead 
of the flux at an observed energy, $E_{obs},$ versus the 
observed flux at 0.3 keV. Here, $\delta_{15}=\delta/15$ 
and $R_{16}=R/10^{16}{\rm cm}$ where $\delta$ and $R$ are 
respectively the source Doppler factor and radius.
The curves labeled (a), (b), (c) are
computed by running a cross-correlation analysis on the lightcurves
shown in Fig 2a, b, and c, respectively, and plotting the lag/lead 
times at which the cross-correlation functions peak. (The cross-correlation
functions are relatively narrow for the lightcurves of Fig. 2.) 
\label{fig3}}

\newpage
\centerline{\psfig{file=fig1.ps}}
\centerline{Figure 1}
\newpage
\centerline{\psfig{file=fig2.ps,height=20.0cm,bbllx=-100pt,bblly=170pt,bburx=546pt,bbury=690pt}}
\centerline{Figure 2}
\newpage
\centerline{\psfig{file=fig3.ps,height=20cm,bbllx=18pt,bblly=44pt,bburx=592pt,bbury=618pt}}
\centerline{Figure 3}

\end{document}